\documentclass{edp-conf}
\usepackage{graphicx}

\newcommand{\Tdust}{T_\mathrm{dust}}
\newcommand{\Xin}{X_\mathrm{in}}
\newcommand{\Xout}{X_\mathrm{out}}

\begin{document}

\TitreGlobal{SF2A 2003}

\title{Formaldehyde emission from low mass protostars}

\author{Maret, S.}\address{Centre d'Etude Spatiale des Rayonnements, 9 avenue du Colonel Roche, BP4346, 31028 Toulouse Cedex 04, France}

\runningtitle{Formaldehyde emission from low mass protostars}

\setcounter{page}{237}

\index{Maret, S.}

\maketitle

\begin{abstract}
  We present a survey of the formaldehyde emission in nine class 0
  protostars obtained with the IRAM 30m and the JCMT millimeter
  telescopes. Using a detailed radiative transfer code of the
  envelopes surrounding the protostars, we show that all but one of
  the observed objects show an inner warm evaporation region where the
  formaldehyde is much more abundant (up to three orders of magnitude)
  than in the outer cold part. The largest inner formaldehyde
  abundances are associated with the sources having the lowest
  submillimetric to bolometric luminosity ratio, i.e. with sources
  closer to the class I border. These abundances are compared with
  predictions from recent models of hot core chemistry.
\end{abstract}
%
\section{Introduction}
 
Formaldehyde is, after water and carbon monoxyde, one of the main
component of ices in grains mantles. Recently, Ceccarelli et
al. (2000a,b), Maret et al. (2002) and Sch{\"o}ier et al (2002) have
shown that in the inner parts of protostellar envelopes, grains
mantles evaporate, releasing the ices components into the gas phase,
and, among them, formaldehyde. Observations of formaldehyde
transitions can be therefore used to determine the physical and
chemical conditions, namely density, temperature and chemical
abundances, in the inner part of protostellar envelopes (Ceccarelli et
al. 2003, Maret et al. 2003). 

The most accepted scenario predict that formaldehyde is formed on
grain surfaces, by successive hydrogenation of CO. The measure of the
formaldehyde abundance in the gaseous phases of the inner part of the
envelopes gives some hints on the composition of the grain mantles, and in
turn on the grain surface chemistry. Moreover, chemistry models predict
that, once in the gas phase, formaldehyde can rapidly form
complex molecules, by the so called \emph{hot core} chemistry
(Charnley et al. 1992). This chemistry was thought to exist only in
high mass protostars, where the gas temperature and density are high
enough to trigger endothermic reactions between species. The very
recent detection of O and N bearing complexes molecules towards
IRAS16293-2422, typical of massive hot cores (Cazaux et al. 2003),
emphasizes the chemical similarity that may exist between low and high
mass protostars.

In order to determine if IRAS16293-2422 is representative of low mass
protostars, or rather a peculiar case, one needs to measure the
formaldehyde abundance in a larger sample of protostars.  In this
contribution, we present the results of a survey of the formaldehyde
emission towards a sample of low mass, Class 0 protostars.

\section{Observations}

A sample of eight Class 0 protostars, located in the Perseus,
$\rho$-Ophiuchus and Taurus complexes, were observed using the James
Clerk Maxwell Telescope and the Institut de Radio Astronomie 30 meter
telescope. Eight formaldehyde transitions were selected, three ortho
and five para, ranging from 140 to 364 GHz. The transitions between
140 GHz and 280 GHz were observed with the IRAM-30m telescope, while
transitions at higher frequencies were observed using the JCMT. The
choice of these two instruments allow a nearly constant beam size over
frequencies. The transitions were chosen to cover a large range of
upper energies, from 20 to 100 cm$^{-1}$, in order to probe the
physical conditions in the different part of the envelope. The
corresponding H$_2^{13}$CO transitions were also observed to determine
the opacity of the lines.

\begin{figure}[h]
  \centering
  \includegraphics[width=9cm]{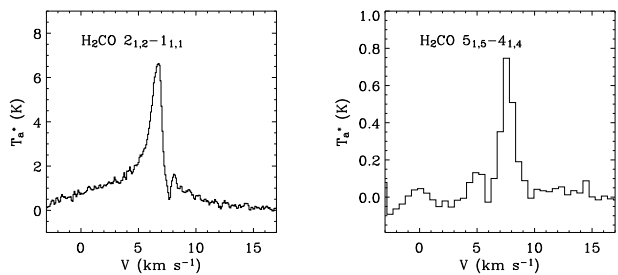}
  \caption{Typical exemple of formaldehyde lines observed towards
  NGC1333-IRAS4A}
  \label{fig:spectrum}
\end{figure}

Fig. \ref{fig:spectrum} shows a typical example of lines observed towards
NGC1333-IRAS4A. The lines are relatively narrow, with a contribution
of the wings extending at larger velocities. Some of the lines show
self-absorption and/or absorption by the foreground material.

\section{Model}

The formaldehyde emission was modeled using a 1D spherical radiative
code. The density and dust temperature profiles determined by
J{\o}rgensen (2002), from the simultaneous modeling of the continuum
emission at 450 and 850 $\mu$m and the spectral energy distribution,
were adopted. The gas temperature was computed by solving the thermal
balance in the envelope (Ceccarelli et al. 1996).

\begin{figure}[h]
  \centering
  \includegraphics[width=9cm]{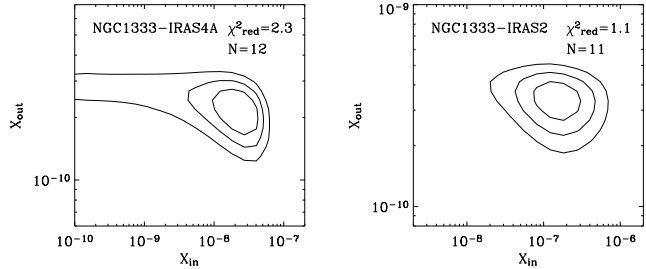}
  \caption{$\chi^{2}$ as a function of $\Xin$ and $\Xout$ for
  NGC1333-IRAS4A and NGC1333-IRAS2. The contour levels show the 1, 2
    and 3 $\sigma$ confidence level respectively.}
  \label{fig:chi2}
\end{figure}

Finally, because of the importance of evaporation in the inner parts
of the envelope, the formaldehyde abundance has been approximated by a
step function: $\Xout$ in the outer part of the envelope where the
dust temperature $\Tdust$ is lower than 100 K, and $\Xin$ in the inner
parts of the envelope where $\Tdust > 100 \mathrm{K}$. These
abundances have been determined by a $\chi^2$ analysis.
Fig. \ref{fig:chi2} show the $\chi^2$ contours has a function of
$\Xin$ and $\Xout$ for two sources of the sample. Table
\ref{tab:xin_xout} presents the derived abundances in all sources.

\begin{table}
  \begin{center}
    \caption{Derived formaldehyde abundances, radius and density where
    $T_\mathrm{dust} = 100 \, \mathrm{K}$ and $L_\mathrm{smm} /
    L_\mathrm{bol}$ ratio.}
    \label{tab:xin_xout}
    \begin{tabular}{l l l l l l l}
      \\
      \hline
      \hline
      Source & $L_\mathrm{smm} / L_\mathrm{bol}$$^\mathrm{a}$ &
      R$_\mathrm{100 K}$ & $n_\mathrm{100 K}$ & $\Xout$ & $\Xin$\\
      & (\%) & (AU) & (cm$^{-3}$) & & \\
      \hline
      NGC1333-IRAS4A & 5 & 53 & 2 $\times$ 10$^{9}$ & 2
      $\times$ 10$^{-10}$ & 2 $\times$ 10$^{-8}$\\
      NGC1333-IRAS4B & 3 & 27 & 2 $\times$ 10$^{8}$ & 5
      $\times$ 10$^{-10}$ & 3 $\times$ 10$^{-6}$\\
      NGC1333-IRAS2 & $\leq 1$ & 47 & 3 $\times$ 10$^{8}$ & 3
      $\times$ 10$^{-10}$ & 2 $\times$ 10$^{-7}$\\
      L1448-MM & 2 & 20 & 2 $\times$ 10$^{8}$ & 7 $\times$
      10$^{-10}$ & 6 $\times$ 10$^{-7}$\\
      L1448-N & 3 & 20 & 1 $\times$ 10$^{8}$ & 3 $\times$
      10$^{-10}$ & 1 $\times$ 10$^{-6}$\\
      L1157-MM & 5 & 40 & 8 $\times$ 10$^{8}$ & 8 $\times$
      10$^{-11}$ & 1 $\times$ 10$^{-8}$\\
      L1527 & 20 & 0.7 & 3 $\times$ 10$^{6}$ & 3 $\times$
      10$^{-10}$ & 6 $\times$ 10$^{-6}$\\
      VLA1623 & 0.2 & 13 & 2 $\times$ 10$^{8}$ & 8 $\times$
      10$^{-10}$ & -\\ 
      \hline
      IRAS16293-2422$^{\mathrm{b}}$ & 2 & 133 & 1 $\times$
      10$^{8}$ & 1 $\times$ 10$^{-9}$ & 1 $\times$ 10$^{-7}$\\
      \hline
    \end{tabular}
  \end{center}
      {\small
        $^{\mathrm{a}}$From Andr\'e et al. (2000).\\
        $^{\mathrm{b}}$From Ceccarelli et al. (2000b).
      }
\end{table}

In all the sources but VLA1623, the observations are only reproduced
if there is a jump in the formaldehyde abundance, between 2 and 3
orders of magnitude. The position of this jump is not well constrained
by our observations, but is consistent with the radius where grain
mantle evaporates in the inner part of the envelope.

It is interesting to note that the higher inner formaldehyde
abundances are observed in the older, namely having the lowest
$L_\mathrm{smm} / L_\mathrm{bol}$ ratio, border class I source,
L1527. \emph{A contrario} the lowest abundances are observed towards
NGC1333-IRAS4A and B, and no jump is observed on the youngest source
of our sample, VLA1623. This result is certainly surprising, as
chemical models predicts that once on the gas phase, formaldehyde will
be rapidly destroyed by endothermic reactions, on a time scale of
10$^{4}$ yr (Rodgers \& Charnley 2003). On the other hand, if the
$L_\mathrm{smm} / L_\mathrm{bol}$ is not a evolutionary tracer, but a
parameter affected by the initial conditions of the pre-stellar core
from which the protostar forms (Jayawardhana 2001), the differences in
the values of $\Xin$ may simply reflect different efficiencies in the
formation of H$_2$CO. Observations on a larger sample of protostars
are needed to answer this question.

\section{Conclusions}

We presented a survey of the formaldehyde emission of a sample of
class 0 protostars. The data have been modeled with a 1D spherical
radiative transfer code. Our model shows that the formaldehyde
abundance is enhanced between two and three orders of magnitude in the
inner part of the envelope, where the dust temperature reaches 100
K. In this region, the grain mantle evaporates, releasing the ices
components, and among them, formaldehyde. The different abundances
observed from one source to the other may reflect different
efficiencies on the formation of H$_2$CO on grain mantles, but other
observations on a larger sample are needed to answer this question.


\end{document}